\documentclass[useAMS,usenatbib]{mn2e}
\usepackage{graphicx}

\title[Molecular gas and star formation]{Is molecular gas necessary for star formation?}
\author[Glover \& Clark]{Simon~C.~O.~Glover \& Paul~C.~Clark \\
Zentrum f\"ur Astronomie der Universit\"at Heidelberg, Institut f\"ur Theoretische
Astrophysik, Albert-Ueberle-Str.\ 2, 69120 Heidelberg \\
 {\tt email:} sglover@ita.uni-heidelberg.de, pcc@ita.uni-heidelberg.de}

\begin{document}

\maketitle

\begin{abstract}
On galactic scales, the surface density of star formation appears to be well correlated with the surface density of molecular gas. This has lead many authors to suggest that there exists a causal relationship between the chemical state of the gas and its ability to form stars -- in other words, the assumption that the gas must be molecular before star formation can occur. We test this hypothesis by modelling star formation within a dense cloud of gas with properties similar to a small molecular cloud using
a series of different models of the chemistry, ranging from one in which the formation of molecules is not followed and the gas is assumed to remain atomic throughout,
to one that tracks the formation of both H$_{2}$ and CO. We find that presence of molecules in the gas has little effect on the ability of the gas to form stars: star formation can occur just as easily in atomic gas as in molecular gas. At low densities ($< 10^4$ cm$^{-3}$), the gas is able to cool via C$^{+}$ fine-structure emission almost as efficiently as via CO rotational line emission, while at higher densities, the main cooling process involves the transfer of energy from gas to dust, meaning that the presence of molecules is again unimportant. Cooling by H$_2$ is particularly inefficient, accounting for as little as 1 percent of the overall cooling in the cloud. Rather than the chemical makeup, we find that the most important factor controlling the rate of star formation is the ability of the gas to shield itself from the interstellar radiation field. As this is also a prerequisite for the survival of molecules within  the gas, our results support a picture in which molecule formation and the formation of cold gas are both correlated with the column density of the cloud -- and thus its ability to shield itself -- rather than being directly correlated with each other.
\end{abstract}


\section{Introduction}
Essentially all Galactic star formation occurs within dense clouds of molecular gas, known as molecular clouds (MCs).
Furthermore, recent observations have demonstrated that on large scales within local galaxies, there is a surprisingly
close correlation between the surface density of star formation, $\Sigma_{\rm SFR}$, and the surface density of molecular
hydrogen, $\Sigma_{\rm H_{2}}$ \citep[see e.g.][]{wong02,leroy08,bigiel08,bigiel11}. An obvious interpretation of these 
observations is that the formation of stars depends on the presence of molecular gas 
\citep{krumholzmckee05,elmegreen07,kmt09}. However, it is not immediately 
obvious why this should be the case. Although H$_{2}$ can be an important coolant of interstellar gas \citep{gk11}, it is 
effective only at temperatures of a few hundred Kelvin or higher. At the temperatures typical of gas within molecular clouds ($T \sim 10$--20~K), the
H$_{2}$ cooling rate is tiny, and hence H$_{2}$ cannot play a direct role in enabling the gas to cool, undergo gravitational
collapse, and form stars. 

A more promising route by which H$_{2}$ can influence the thermodynamics of the gas is through the fact that its presence is 
required for the efficient formation of carbon monoxide (CO). Unlike H$_{2}$, CO can provide effective cooling even in very low 
temperature gas, and it is generally found to be the dominant gas-phase coolant within prestellar cores \citep{nlm95,gold01}. 
However, even if the H$_{2}$ and CO were absent, the gas would still be able to cool to low temperatures through fine structure emission
from ionized and neutral atomic carbon. In previous modelling, we have shown that C$^{+}$ cooling alone can reduce the
gas temperature to $T \sim 20$~K within gas that is shielded from the effects of photoelectric heating by dust extinctions of
$A_{\rm V} \sim 1$--2 or more \citep{gm07}. To reduce the temperature further, to the 10~K typical of most prestellar cores, 
a molecular coolant such as CO is still required, but it seems unlikely that the difference between the $\sim 10$~K temperatures
reachable with molecular cooling and the $\sim 20$~K temperatures reachable with atomic fine structure cooling can have any
great effect on the ability of gas to form stars. 

One is therefore led to ask whether the presence of molecules  is truly essential for star 
formation. Perhaps the reason that we find such a good correlation between molecular gas and star formation is not that
molecular gas is a prerequisite for star formation, but instead that the formation of molecules and the formation of stars both
correlate with some other factor, such as the existence of large, dense clouds of gas. In this picture, the molecules are a
consequence, not a cause, of the conditions required to form stars.

To help us better understand whether or not the presence of molecular gas is a necessary condition for star formation, we
have performed a number of numerical simulations of the formation of stars in a dense cloud of gas with properties similar
to a small molecular cloud. We use a range of different models of the chemistry and thermal balance of the gas to explore
the influence that molecule formation and dust shielding have on the ability of the cloud to form stars. 
In Section~\ref{numerics} of this paper, we discuss the numerical 
setup that we use to perform these simulations, and our choice of initial conditions. The results are presented in 
Section~\ref{results}, and in Section~\ref{discuss} we discuss what these results imply for the nature of the link between 
molecular gas and star formation. We conclude in Section~\ref{conc}.

\section{Simulations}
\label{numerics}
\subsection{Numerical approach}
Our simulations were performed using a modified version of the Gadget 2 smoothed particle hydrodynamics (SPH) 
code \citep{springel05}. Our modifications include a sink particle algorithm for treating gravitationally collapsing regions
that become to small to resolve \citep{bbp95}, the inclusion into the equation of motion of an optional confining pressure
term \citep{benz90}, a treatment of gas-phase chemistry (described in more detail below), as well as radiative heating and 
cooling from a number of atomic and molecular species \citep{gj07,g10,gc11}, and an approximate treatment of the attenuation 
of the Galactic interstellar radiation field (ISRF). The effects of magnetic fields are not included.

In order to establish the role that molecular cooling plays in determining the star formation rate, we have run a series of 
simulations a small, dense gas cloud with properties similar to those of a small molecular cloud, using several different treatments
of the cooling and chemistry of the gas. In our least realistic model (run A), we include the simple network for hydrogen chemistry 
described in \citet{gm07}, together with the model for CO chemistry developed by \citet{nl97}, but we assume that the gas remains 
optically thin throughout the simulation;  i.e.\ we neglect any attenuation of the ISRF by dust absorption or molecular self-shielding. 
In practice, this means that the gas remains in atomic form at all but the highest densities (see Section~\ref{therm} below). In run B, 
we include the effects of dust absorption, but ignore the chemical evolution of the
gas, forcing it to remain in atomic form throughout. To model the dust absorption, we use our new {\em Treecol} algorithm,
as described in more detail in Section~\ref{treecol} below. In run C, we include the simple network for hydrogen chemistry 
described in \citet{gm07}, but assume that the carbon and oxygen in the gas remain as C$^{+}$ or O, respectively. In this model, 
the only molecular cooling comes from H$_{2}$. Finally, in runs D1, D2 and D3, we include both the \citet{gm07} hydrogen 
chemistry and the \citet{nl97} CO chemistry, this time including the effects of dust absorption and self-shielding. This setup 
has been shown to produce more CO than would be the case with more detailed chemical treatments \citep{gc11}, and hence 
gives us a reasonable upper bound on the effectiveness of CO cooling. We performed three simulations using this treatment of the 
chemistry: one in which the hydrogen is initially atomic (run D1), and two in which it is initially molecular (runs D2, D3). These
final two runs differ only in the numerical resolution adopted: run D2 is performed using our standard resolution (discussed
in Section~\ref{ic} below), while run D3 is performed using ten times more SPH particles, and hence has ten times better mass
resolution. A brief summary of these models is given in Table~\ref{sims}.

None of the models that we study in this paper include the effects of freeze-out of species such as CO onto dust grains. 
However, we do not expect this simplification to significantly affect the thermal balance of the cloud \citep[see e.g.][]{gold01}.

\begin{table}
\caption{Simulation details \label{sims}}
\begin{tabular}{lcccc}
\hline
& \multicolumn{2}{c}{Chemistry} & & \\
ID & H$_{2}$ & CO & Initial state & Notes \\
\hline
A & Y & Y & Atomic & Optically thin \\
B & N & N & Atomic & \\
C & Y & N & Atomic & \\
D1 & Y & Y & Atomic & \\
D2 & Y & Y & Molecular & \\
D3 & Y & Y & Molecular & High resolution \\
\hline
\end{tabular}
\end{table}

\begin{figure}
\includegraphics[width=3.5in]{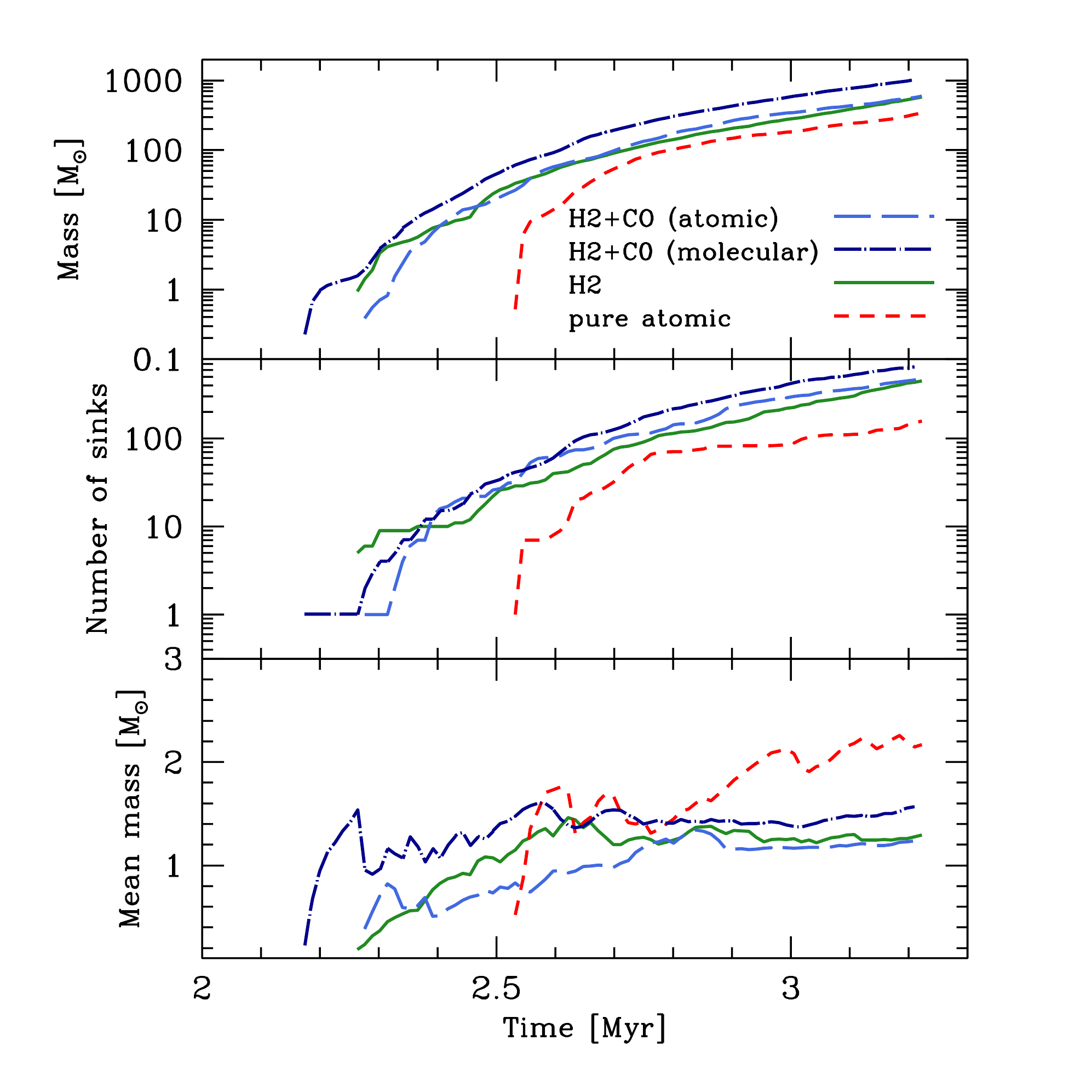}
\caption{{\em Upper panel:} mass in sinks as a function of time in runs B (shorted-dashed line), C (solid line),
D1 (long-dashed line) and D2 (dot-dashed line). {\em Middle panel:} number of sinks formed in each simulation,
plotted as a function of time. {\em Lower panel:} mean sink mass versus time in the same four simulations. For
reference, the mass resolution in these simulations was $0.5 \: {\rm M_{\odot}}$. \label{SF-hist}}
\end{figure}

\subsection{Attenuation of the ISRF}
\label{treecol}
We assume that our model clouds are illuminated by the standard interstellar radiation field, using the parameterization
from \citet{dr78} in the UV, and from \citet{bl94} at longer wavelengths. To model the effects of extinction, H$_{2}$ self-shielding, 
CO self-shielding and the shielding of CO by H$_{2}$, we use the {\em Treecol} algorithm (Clark et~al., in prep.)
This  algorithm allows us to approximate the distribution
of column densities (of hydrogen nuclei, molecular hydrogen or CO) seen by each particle by using information stored in the 
oct-tree that is also used in the Gadget 2 code to compute gravitational accelerations. Our current implementation of {\em Treecol}
uses the Healpix pixelation scheme \citep{healpix} and yields for each SPH particle a 4$\pi$ steradian projection of the
column density distribution seen by that particle, discretized into 48 equal-area Healpix pixels.

To illustrate how we can use this pixelated projection to compute photoheating or photodissociation rates for a given 
SPH particle, let us follow how we would go about computing the photoelectric heating rate. We begin with 
the appropriate pixelated column density distribution, which in this case is that for the column density of hydrogen nuclei in 
all forms, $N_{\rm H, tot}$. We use the relationship $A_{\rm V} = 5.348 \times 10^{-22} (N_{\rm H, tot} / 1 \: {\rm cm^{-2}})$ 
between $N_{\rm H, tot}$ and  the visual extinction $A_{\rm V}$  \citep{bsd78,db96} to associate a visual extinction with each pixel. 
We next compute an attenuation factor $f_{\rm pe}$ for 
each pixel, which we define as $f_{\rm pe} = \Gamma_{\rm pe}(A_{\rm V}) / \Gamma_{\rm pe}(0)$, where $\Gamma_{\rm pe}(A_{\rm V})$ 
is the photoelectric heating rate in gas with a visual extinction equal to $A_{\rm V}$, and $\Gamma_{\rm pe}(0)$ is the
photoelectric heating rate in the optically thin limit. We take our expressions for these quantities from \citet{berg04}.
Finally, once we have computed $f_{\rm pe}$ for each pixel, we make use of the fact that the Healpix pixels have equal areas
to compute a mean attenuation factor $\langle f_{\rm pe} \rangle$ for the gas represented by the SPH particle by averaging
$f_{\rm pe}$ over all pixels. The photoelectric heating rate in this gas is then just $\Gamma_{\rm pe} = \langle f_{\rm pe} \rangle 
\Gamma_{\rm pe}(0)$.

We use the same strategy to compute the rate at which dust is heated by the interstellar radiation field, taking our 
expression for the dependence of the heating rate on $A_{\rm V}$ from \citet{gc11}. We also use a similar strategy to 
compute the H$_{2}$ and CO photodissociation rates. However, in this case it is necessary to
use the {\em Treecol} algorithm to compute the
distributions of H$_{2}$ and CO column densities in addition to the total hydrogen column density, since these are 
needed in order to compute the mean attenuation factors corresponding to H$_{2}$ self-shielding, $\langle f_{\rm H_{2}, H_{2}} 
\rangle$, CO self-shielding, $\langle f_{\rm CO, CO} \rangle$, and the shielding of CO by H$_{2}$, $\langle f_{\rm H_{2}, CO} \rangle$.
Our expression for $f_{\rm H_{2}, H_{2}}$ comes from \citet{db96}, while those for $f_{\rm H_{2}, CO}$ and $f_{\rm CO, CO}$ come
from \citet{lee96}.  A more accurate treatment of CO self-shielding and the shielding of CO by H$_{2}$ has recently been given 
by \citet{visser09}, but we would not expect our results to change significantly if we were to use this in place of the older \citet{lee96}
treatment.

\begin{figure}
\includegraphics[width=3.1in]{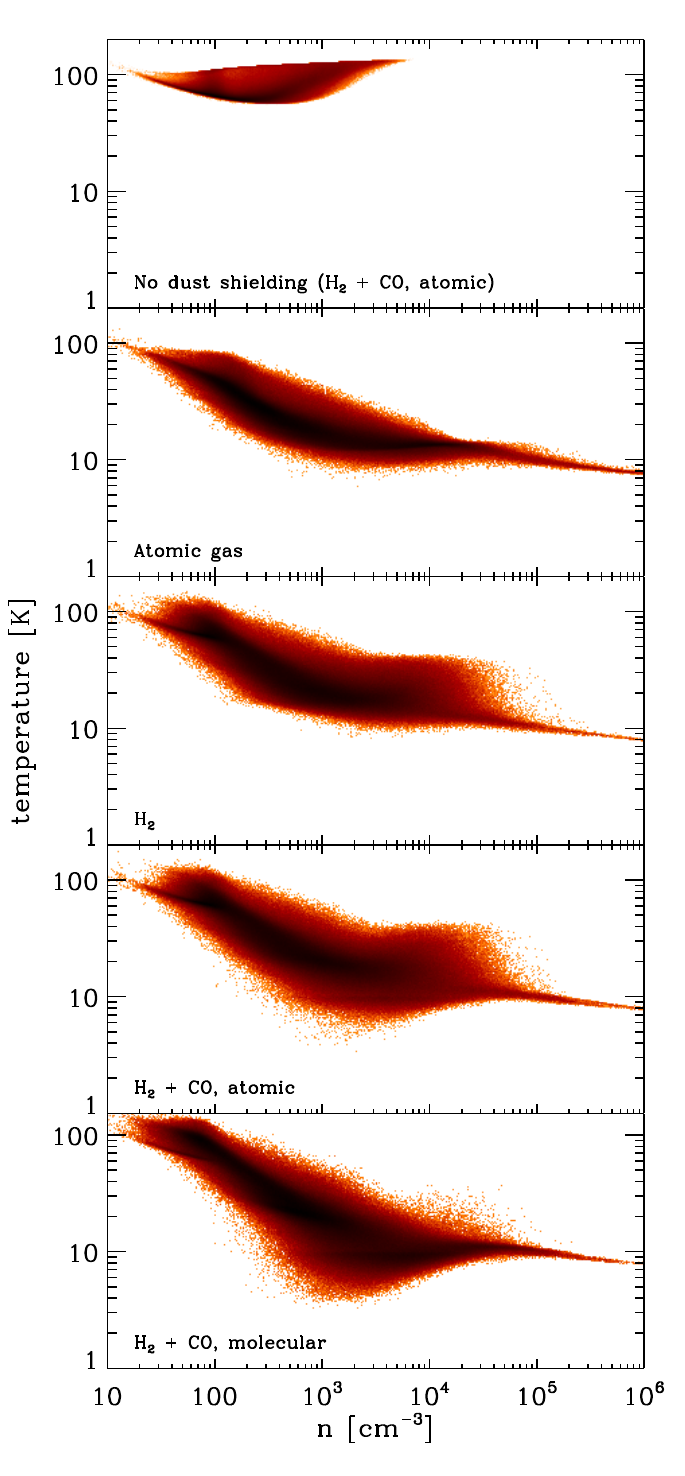}
\caption{Gas temperature plotted as a function of $n$, the number density of hydrogen nuclei,
in runs B, C, D1 and D2 (panels 2--5) at a time immediately prior to the onset of star formation in each of 
these runs. Note that this means that each panel corresponds to a slightly different physical
time. For comparison, we also plot the temperature of the gas as a function of density in run
A (panel 1, at the top) at a similar physical time, $t = 2.3 \: {\rm Myr}$, although in this case, this is long
before the cloud begins to form stars.
\label{tphase}}
\end{figure}

\subsection{Sink particles}
Sink particles are formed once the gas exceeds a number density of $10^7$ cm$^{-3}$, which is around the density at which we start to lose resolution in the $2 \times 10^6$ particle simulations (assuming a minimum temperature of 8K; see Section~\ref{therm} below). To ensure that sink particles are not created by mistake, we require the parcel of gas comprising the candidate particle and its 50 nearest neighbours to be gravitationally bound and collapsing (following the prescription given by \citealt{bbp95}). Once formed, sink particles are able to accrete additional SPH particles coming within a preset sink accretion radius $r_{\rm acc}$, provided that the
SPH particles are not only gravitationally bound to the sink, but are also more strongly bound to it than to any other sink within the computational volume. We prevent
new sink particles from forming within a distance of $2\,r_{\rm acc}$ of existing sink particles.  In this study, we set $r_{\rm acc} = 0.0026 \: {\rm pc}$,  equal to the Jeans length at $n = 10^7$ cm$^{-3}$ and T = 8~K when the mean molecular weight $\mu = 2.33$ amu. All gravitational forces, including those from the sinks,
are softened using the standard kernel softening that has been shown by a number of authors to prevent `artificial' fragmentation of the gas \citep{bb97, w98, hgw06}.

\subsection{Initial conditions}
\label{ic}

The starting point for our calculations is a uniform sphere with a hydrogen nuclei number density $n = 300 \: {\rm cm}^{-3}$, and a total mass of 10$^4 \rm{M}_\odot$, 
meaning that it has an initial radius of approximately 6~pc. Such conditions are typical of those found in nearby star-forming clouds, and give the cloud a  visual
extinction at its centre of roughly three.
We model the cloud with either $2 \times 10^6$ SPH particles (our standard resolution) or $2 \times 10^7$ SPH particles (our high resolution, used only in run D3), and we require that every SPH particle's quantities are calculated by averaging over 50 neighbours. This sets the mass resolutions at 0.5 and 0.05 $\rm{M}_\odot$ in our low and high resolution simulations respectively \citep{bb97}. We inject bulk (non-thermal) motions into the cloud by imposing a turbulent velocity field that has a power spectrum of $P(k) \propto k^{-4}$. The energy in the turbulence is initially equal to the gravitational energy in the cloud, giving an initial root mean squared velocity of around 3 km\,s$^{-1}$. The turbulence is left to freely decay via shocks and compression-triggered cooling. We set the initial temperature of the gas to 20K, but this is quickly altered when the simulation starts, as the gas tries to find an equilibrium between the various heating and cooling processes that are included in our model of the interstellar medium. We adopt a confining pressure equal to $p_{\rm ext} = 12000 \: {\rm K} \: {\rm cm^{-3}}$, but note that runs with much smaller values of 
$p_{\rm ext}$ produce very similar results.

We adopt total carbon and oxygen abundances relative to hydrogen that are $x_{\rm C} = 1.4 \times 10^{-4}$ and
$x_{\rm O} = 3.2 \times 10^{-4}$, respectively \citep{sem00}. For the cosmic ray ionization rate of atomic hydrogen, we take 
a value $\zeta_{\rm H} =  10^{-17} \: {\rm s^{-1}}$. Cosmic ray ionization rates for other species (e.g.\ molecular hydrogen) 
were computed assuming that they had the same ratios to $\zeta_{\rm H}$ as given in the UMIST99 chemical database
\citep{teu00}.

\section{Results}
\label{results}
\subsection{Star formation in the clouds}
In the top panel of Figure~\ref{SF-hist}, we show the rate at which mass is converted into sink
particles in runs B, C, D1 and D2. In all four of these runs, star formation begins within the cloud
after roughly 2 -- 2.5~Myr. For comparison, the global free-fall time of the cloud in its initial state
is approximately 2.5 Myr, and so in all four of these runs, stars begin to form within roughly one
free-fall time.

If we look in more detail at the time histories plotted in Figure~\ref{SF-hist}, we see that the time that
elapses between the beginning of the simulations and the onset of star formation has a slight
sensitivity to the chemical state of the gas. Run D2, which includes both H$_{2}$ 
and CO chemistry, and which begins with the hydrogen already fully molecular, is the first run to 
begin forming stars. However, star formation in run C (initially atomic, H$_{2}$ chemistry only)
and run D1 (initially atomic, H$_{2}$ and CO chemistry) is delayed only slightly compared 
to run D2, by roughly 0.1~Myr. Moreover, by the end of the simulation, all three simulations are 
forming stars at very nearly the same rate, although a greater number of stars have formed in
run D2, owing to its head start. Finally, star formation in run B (atomic gas) is delayed for slightly 
longer, roughly 0.3~Myr, although once star formation does begin, it occurs at a very similar rate
to that in the other three runs. 

Nevertheless, the differences between the star formation histories of the clouds simulated in these
four runs are relatively small, despite the significant differences that exist in the chemical make-up
of the clouds. The presence of H$_{2}$ and CO within the gas appears to make only a small difference
in the ability of the cloud to form stars. Furthermore, the fact that a cloud that does not form {\em any} molecules
is not only able to form stars, but does so with only a short delay compared to one in which all of the 
hydrogen and a significant fraction of the carbon is molecular is persuasive evidence that the 
formation of molecules is not a prerequisite for the formation of stars.

Examination of the nature of the stars formed in these four simulations is also informative. Naively,
one might expect that in the absence of molecular cooling, or more specifically CO cooling, the 
minimum temperature reached by the star-forming gas would be significantly higher. If so, then 
this would imply that the value of the Jeans mass in gas at this minimum temperature would also be
significantly higher.  It has been argued by a number of authors \citep[e.g.][]{larson05,jappsen05,bonnell06} 
that it is the value of the 
Jeans mass at the minimum gas temperature in a star-forming cloud that determines the characteristic 
mass in the resulting initial mass function (IMF). Following this line of argument, one might therefore expect
the characteristic mass to be much higher in clouds without CO. However, our results suggest that this
is not the case. In runs C, D1 and D2, the mean mass of the stars that form is roughly 1 -- 1.5 M$_{\odot}$
(Figure~\ref{SF-hist}, bottom panel) and shows no clear dependence on the CO content of the gas.
Indeed, the mean mass is slightly lower in run C, which has no CO, than in run D2, which does.
The mean stellar mass that we obtain from these simulations is slightly higher than the characteristic
mass in the observationally-determined IMF, which is typically found to be somewhat less than a solar
mass \citep{chabrier01,kroupa02}. However, this is
a consequence of our limited mass resolution, which prevents us from forming stars less massive
than 0.5 M$_{\odot}$, and hence biases our mean mass towards higher values. (We return to this point
in Section~\ref{resn} below). In run B, we again find a greater difference in behaviour, but even in this case, 
the mean stellar mass remains relatively small, at roughly 2 M$_{\odot}$. It therefore appears that the presence 
or absence of molecules does not strongly affect either the star formation rate of the clouds or the mass function 
of the stars that form within them. 

Conspicuous by its absence from our discussion so far has been run A, the run in which we assumed that 
the gas remained optically thin throughout the simulation. In this run, we find a very different outcome. Star
formation is strongly suppressed, and the first star does not form until $t = 7.9$~Myr, or roughly three global
free-fall times after the beginning of the simulation. The results of this run suggest that it is the ability of the
cloud to shield itself from the effects of the interstellar radiation field, rather than the formation of molecules
within the cloud, that plays the most important role in regulating star formation within the cloud. 
\citep[c.f.][]{klm11}

\subsection{Thermal and chemical state of the gas}
\label{therm}
\subsubsection{Temperature distribution}
In order to understand why molecular gas appears to be of only very limited importance in determining the star
formation rate, it is useful to look at the thermal  state of the gas in the different runs at the point at
which they begin forming stars. This is illustrated in Figure~\ref{tphase} for runs B, C, D1 and D2. For comparison,
we also show the temperature distribution of the gas in run A at $t = 2.3 \: {\rm Myr}$ (i.e.\ at a similar time to the
other four runs, albeit roughly 5.6~Myr before run A itself begins to form stars). The first point to note is the basic 
similarity of the temperature distribution in most of the 
runs. In runs B, C, D1 and D2, the temperature decreases from roughly 100~K at $n \sim 10 \: {\rm cm^{-3}}$ to 10~K at 
$n = 10^{5} \: {\rm cm^{-3}}$ and 8~K at $n = 10^{6} \: {\rm cm^{-3}}$. This corresponds to a relationship between
temperature and density that can be approximated as $T \propto \rho^{-0.25}$ at $n < 10^{5} \: {\rm cm^{-3}}$,
or a relationship between pressure and density $P \propto \rho^{0.75}$, in good agreement with the relationship
$P \propto \rho^{0.73}$ proposed by \citet{larson85,larson05}. The fact that the effective equation of state of the
gas is significantly softer than isothermal (i.e.\ $P \propto \rho$) means that the local Jeans mass decreases
rapidly with increasing density within the cloud, a factor which is known to greatly assist gravitational fragmentation
\citep[see e.g.][]{li03,jappsen05,clark08,dopcke11}. In run A, on the other hand, there is a brief period of cooling at low densities,
but the gas then begins to reheat at densities $n > 500 \: {\rm cm^{-3}}$, and the temperature in the densest gas is
higher than the temperature at the edge of the cloud. It is therefore not surprising that gravitational fragmentation is
strongly suppressed in this run.

If we look in more detail at the relationship between temperature and density in the four runs that do show significant
cooling, then several additional features become apparent. First, all of the runs show a significant degree of scatter
in the relationship at densities between $n \sim 100 \: {\rm cm^{-3}}$ and $n \sim 10^{4} \: {\rm cm^{-3}}$ (in run B) or
$n \sim 10^{5} \: {\rm cm^{-3}}$ (in the other runs), but this scatter abruptly
vanishes at higher densities. This pronounced change in behaviour is due in part to a change in the dominant coolant within
the clouds. At low densities, gas-phase coolants (primarily C$^{+}$ and CO) dominate, 
while at higher densities, energy transfer between gas and dust becomes the dominant process regulating the gas
temperature. In the regime dominated by gas-phase cooling, the local cooling rate depends on the local value of
the optical depth in the appropriate cooling line, and hence on the local details of the velocity structure of the cloud.
In the dust-dominated regime, on the other hand, the gas temperature quickly converges to the dust temperature, 
which is determined largely by continuum emission and absorption and which is thus insensitive to the velocity
structure of the cloud. In addition, another effect contributing to the scatter at low densities is the sensitivity of the
photoelectric heating rate to the visual extinction. We know from previous work that the mean visual extinction 
within a given fluid element in a turbulent cloud is only poorly correlated with the volume density of the cloud
\citep{g10}, and hence the value of the photoelectric heating rate shows a significant scatter at any given density.
At low densities, photoelectric heating is the dominant heat source (see Section~\ref{dom_therm} below), and so this scatter in
the heating rate helps to create a significant scatter in the gas temperature. At higher densities, photoelectric heating
becomes far less important, and so the scatter in the photoelectric heating rate has much less effect on the 
temperature.

A second feature to note is the pronounced ``hump'' in the temperature-density distribution, roughly
centered at $T \sim 20$~K and $n \sim 2 \times 10^{4} \: {\rm cm^{-3}}$, that is present in runs C and 
D1, but absent in runs B and D2. This feature is caused by H$_{2}$ formation heating: we assume \citep[following][]{tu01}
that H$_{2}$ molecules form on dust grains with a high degree of rotational and vibrational excitation,
and at densities above a few thousand particles per cubic centimetre, much of this energy is converted into thermal
energy by collisional de-excitation of the newly-formed H$_{2}$ molecules. This does not occur in run B
because no H$_{2}$ is allowed to form in that run, and it does not occur in run D2 because the hydrogen starts
in fully molecular form, and although some is subsequently dissociated, the atomic hydrogen fraction at the
appropriate densities remains small (see Figure~\ref{rhoh2}).

Finally, the influence of the CO in runs D1 and D2 is clear if we look at the temperature distribution of the gas
at densities of around $10^{3}$ to $10^{4} \: {\rm cm^{-3}}$. The minimum gas temperature reached in the runs
with CO is approximately 5~K, roughly half the size of the minimum temperature reached in the runs that must rely 
on C$^{+}$ cooling. However, despite the fact that the gas is able to reach lower temperatures in the runs with CO
cooling, very little of it does so: most of the gas at these densities has a temperature that is close to 10~K, implying
that the difference in the mean temperature of the gas is much smaller than the difference in the minimum 
temperature.

Given the similarity between the temperature distribution in run B and that in runs C, D1 and D2, it may at first
seem odd that star formation in the former is delayed by roughly 0.3~Myr compared to star formation in the 
latter, and that the mean mass of the stars that form is noticeably larger. However, this has a simple explanation.
At a given density and temperature, the Jeans mass scales with the mean molecular weight $\mu$ as
$M_{J} \propto \mu^{-3/2}$. In run B, the gas is fully atomic at all densities, and so $\mu = 1.27 \: {\rm amu}$ (where we 
have assumed a helium fraction of 0.1 by number, relative to hydrogen). On the other hand, in runs C, D1 and D2,
the hydrogen in the densest gas is almost fully molecular, as illustrated in Figure~\ref{rhoh2}, 
and so $\mu = 2.33 \:  {\rm amu}$.  Therefore, for any given choice of density and temperature within the dense
gas, the Jeans mass is roughly a factor of 2.5 times larger in run B than in runs C, D1 or D2. If we assume that
gas is assembled into dense filaments and cores at roughly the same rate in all of these runs, which seems likely
given the similar temperature-density relationship, then this implies
that in run B it will take longer for these structures to grow to the point at which they become gravitationally unstable,
and hence that the onset of star formation will be delayed. It also suggests that the stars that do form will 
have systematically larger masses, in agreement with what we find in our simulations.

\subsubsection{Density distribution}
\begin{figure}
\includegraphics[width=3.3in]{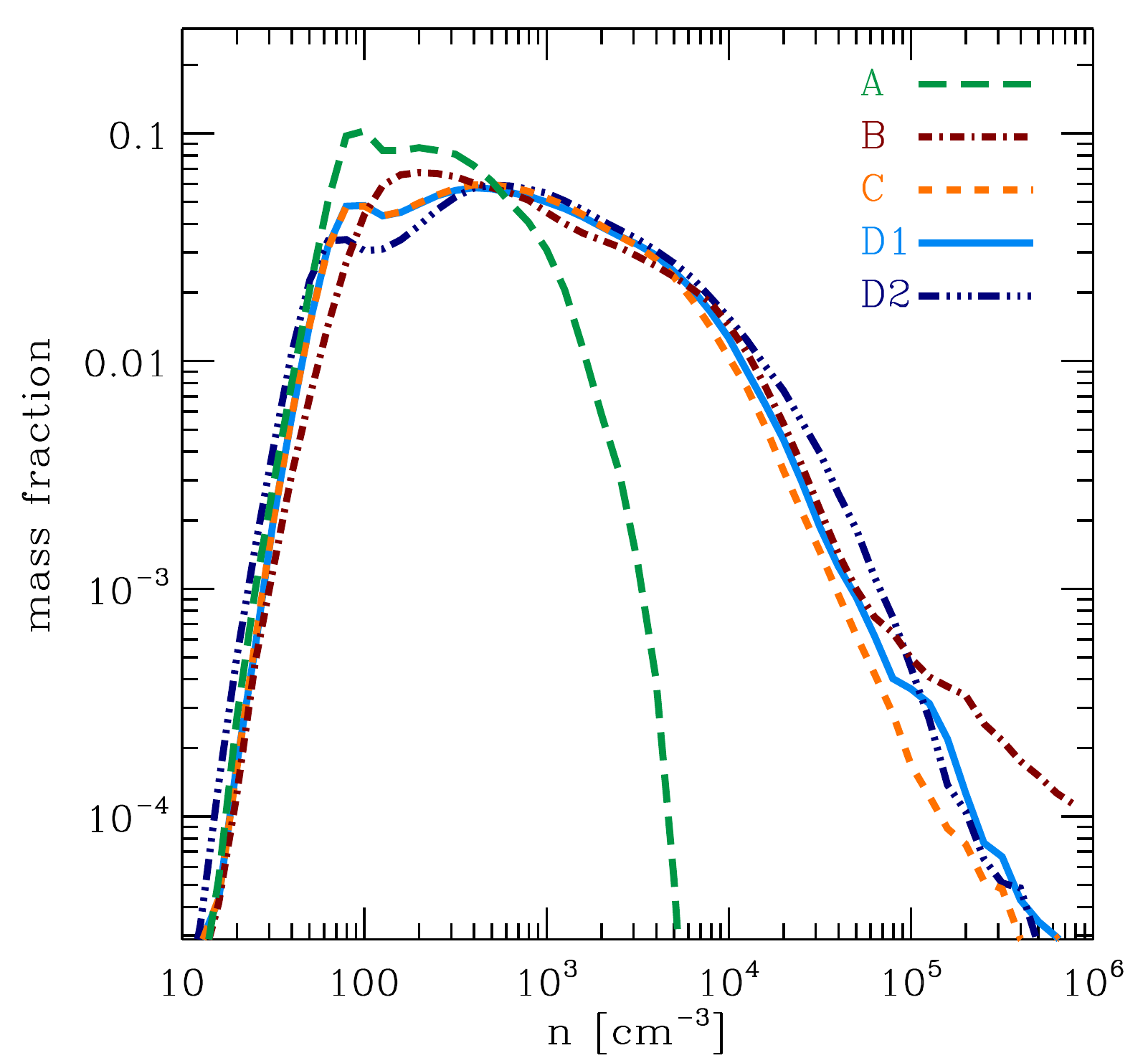}
\caption{Comparison of the mass-weighted density PDF in the various runs.  For runs B, C, D1 and D2, 
we show the state of the gas immediately prior to the onset of star formation, while for run A, we show 
the state of the gas at a similar physical time, $t = 2.3 \: {\rm Myr}$.
\label{rhopdf}}
\end{figure}

We can gain additional insight into the physical processes determining the star formation rate in our model clouds
by examining the probability density function (PDF) of gas density generated in the different runs. In Figure~\ref{rhopdf},
we show the mass-weighted density PDF of the gas in runs B, C, D1 and D2 immediately prior to the onset of star
formation in these four runs. For comparison, we also show the density PDF in run A at a similar physical time,
$t = 2.3 \: {\rm Myr}$. 

In the four star-forming runs, the density PDF is broad, and shows evidence for a power-law
tail at high densities, $n > 10^{4} \: {\rm cm^{-3}}$. Similar features have been found in previous simulations of
self-gravitating supersonic turbulence \citep[e.g.][]{klessen00,fed08,knw11}, in which run-away gravitational collapse 
has occurred. There is also now observational
evidence for the presence of power-law tails in the density PDFs of active star-forming regions in real molecular clouds \citep{kain09,lom10,kain11}.
In contrast, the density PDF in run A is much narrower, and appears to have a log-normal form, as would be expected in 
a cloud in which gravitational collapse has not yet occurred. The higher gas temperatures in run A compared to the other
runs mean that the turbulence has a much smaller root-mean-squared Mach number, and hence produces weaker 
compressions, leading to a significantly narrower PDF  \citep{pnj97,fks08,pn11}. As a result, none of the gas in run A has
yet reached a high enough density to become locally self-gravitating and to decouple from the larger-scale flow, while in the
other four runs, a significant fraction of the gas mass is found in collapsing, self-gravitating structures. 

The fact that the four star-forming runs have very similar density PDFs is a consequence of the fact that they have 
very similar temperature distributions. It is therefore easy to understand why they form stars at very similar rates:
in all four cases, the amount of dense gas available for star formation is very similar, and the rate at which this dense
gas is converted into stars is limited primarily by the local free-fall time, which does not depend upon the nature of
the gas coolants.

\subsubsection{Molecular abundances}
\begin{figure}
\includegraphics[width=3.3in]{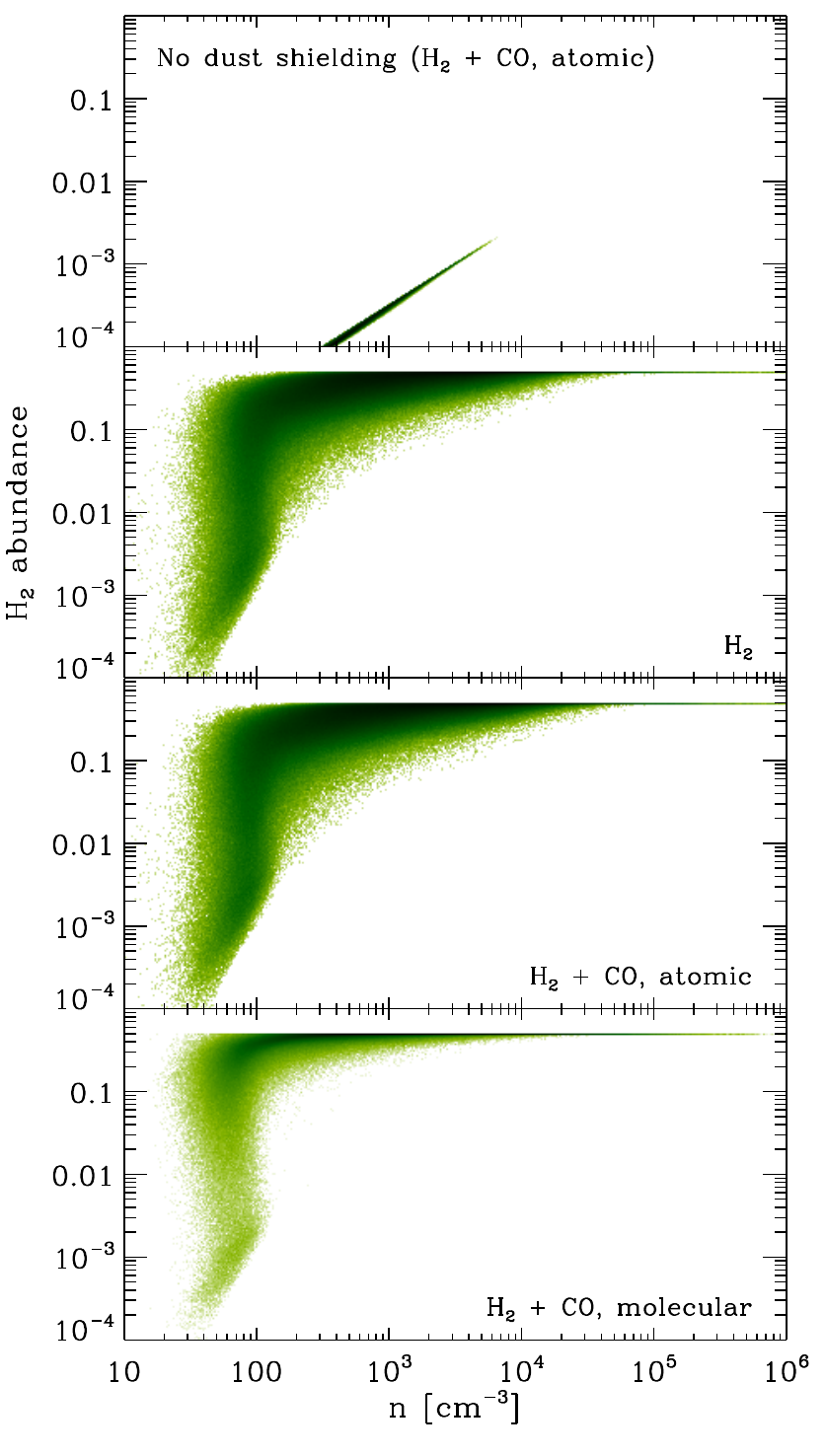}
\caption{Fractional abundance of H$_{2}$ (defined as the ratio of the H$_{2}$ number density
$n_{\rm H_{2}}$ and the number density of hydrogen nuclei, $n$), plotted as a function of $n$,
for runs A, C, D1 and D2. With this definition of the fractional abundance, a value of 0.5 corresponds
to fully molecular gas. For runs C, D1 and D2, we show the state of the gas immediately prior to the
onset of star formation, while for run A, we show the state of the gas at a similar physical time, $t = 2.3 
\: {\rm Myr}$.
\label{rhoh2}}
\end{figure}

In Figure~\ref{rhoh2}, we plot the fractional abundance of H$_{2}$, $x_{\rm H_{2}}$, as a function of the hydrogen 
nuclei number density $n$ in runs C, D1 and D2 at a time shortly before the onset of star formation in these 
simulations. For comparison, we also plot $x_{\rm H_{2}}$ versus $n$ at a similar output time in run A. We show 
no results from run B because $x_{\rm H_{2}} = 0$ by design in this run. We define $x_{\rm H_{2}}$ as the 
ratio of the H$_{2}$ number density $n_{\rm H_{2}}$ and the hydrogen nuclei number density $n$, which means
that in fully molecular gas, $x_{\rm H_{2}} = 0.5 n / n= 0.5$.

If we compare the results from runs C, D1 and D2, we see that in each case, the cloud is predominantly
molecular at the point at which it forms stars. The mean H$_{2}$ fractional abundance falls significantly below 
0.5 only for densities $n < 100 \: {\rm cm^{-3}}$, corresponding to warm, unshielded gas at the boundaries
of the clouds. The rapidity with which H$_{2}$ forms within the clouds is a consequence of the many 
transient density enhancements produced by the supersonic turbulence, which boost the overall H$_{2}$ 
formation rate in the cloud by a factor of a few \citep[for more details, see e.g.][]{gm07,mili11}.
The main difference of note between runs C, D1 and D2 is the larger scatter in $x_{\rm H_{2}}$
at high densities in runs C and D1 than in run D2. This reflects the difference in initial conditions between
these runs: runs C and D1 start with all of their hydrogen in atomic form, and although most of it has become
molecular by $t \sim 2.5 \: {\rm Myr}$ there are still a few regions that retain a high atomic fraction. In run D2,
on the other hand, the hydrogen is initially in fully molecular form, and atomic hydrogen is produced only by
photodissociation. The main conclusion that we can draw from this is that at the point at which star formation
begins, the clouds have not yet reached chemical equilibrium throughout their volume.

In run A, on the other hand, the situation is very different. In the absence of H$_{2}$ self-shielding and
dust shielding, the H$_{2}$ fractional abundance is much smaller: the peak value at the displayed
output time of $2.3 \: {\rm Myr}$ is $x_{\rm H_{2}} \simeq 2 \times 10^{-3}$, implying that even in the
densest gas, less than 1\% of the hydrogen is in molecular form. In addition, the very small degree of
scatter in the relationship between $x_{\rm H_{2}}$ and $n$ implies that the gas is in chemical 
equilibrium, as we would expect given that in the absence of shielding, the H$_{2}$ photodissociation
timescale $t_{\rm pd} \simeq 600 \: {\rm yr}$, which is very much shorter than the age of the cloud.

\begin{figure}
\includegraphics[width=3.3in]{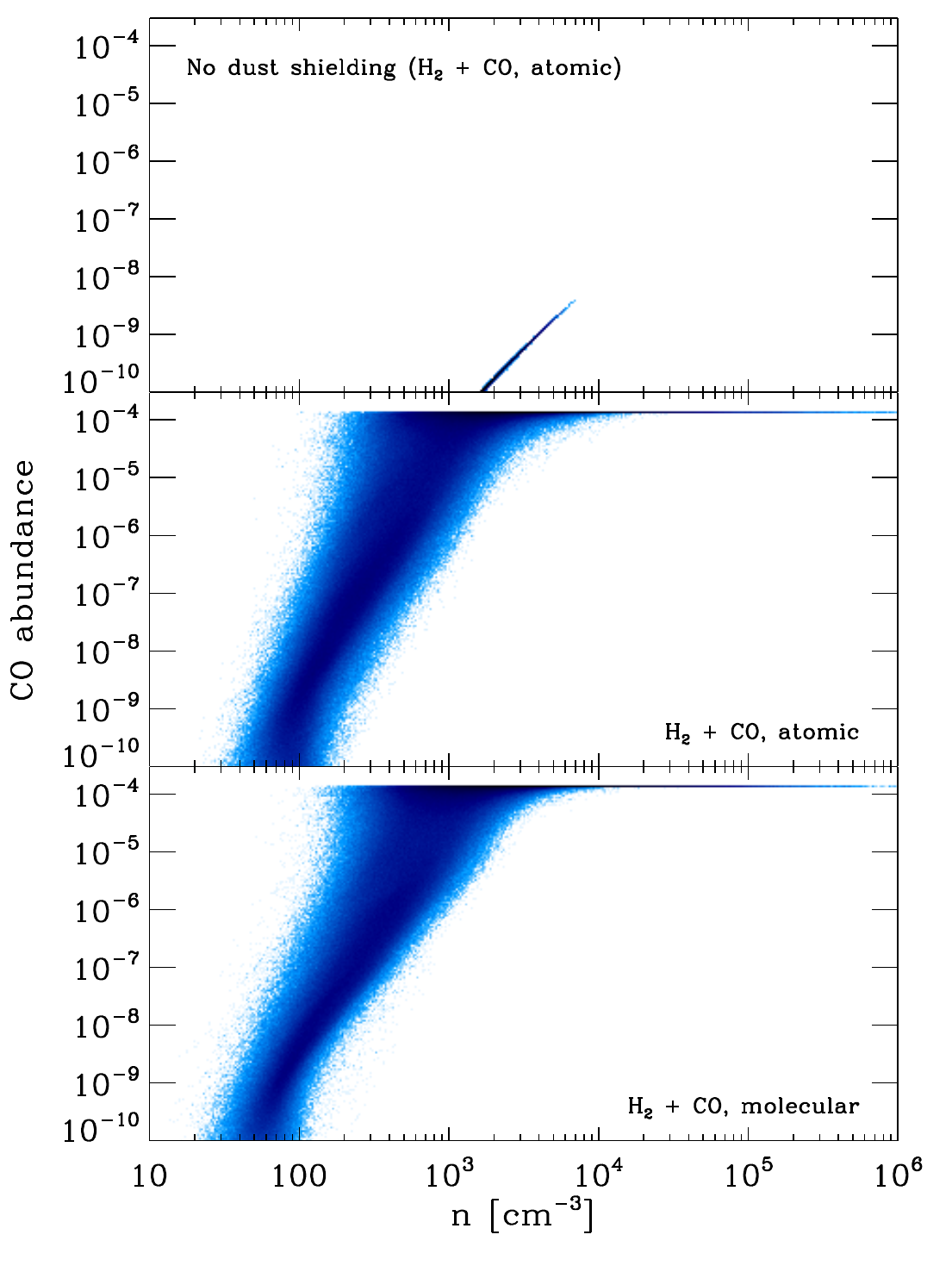}
\caption{As Figure~\ref{rhoh2}, but for the fractional abundance of CO. In this case, results are plotted
only for runs A, D1 and D2, as the CO content of the gas in runs B and C is zero, by design.
\label{rhoco}}
\end{figure}

\begin{figure*}
\centerline{
\includegraphics[width=3.6in]{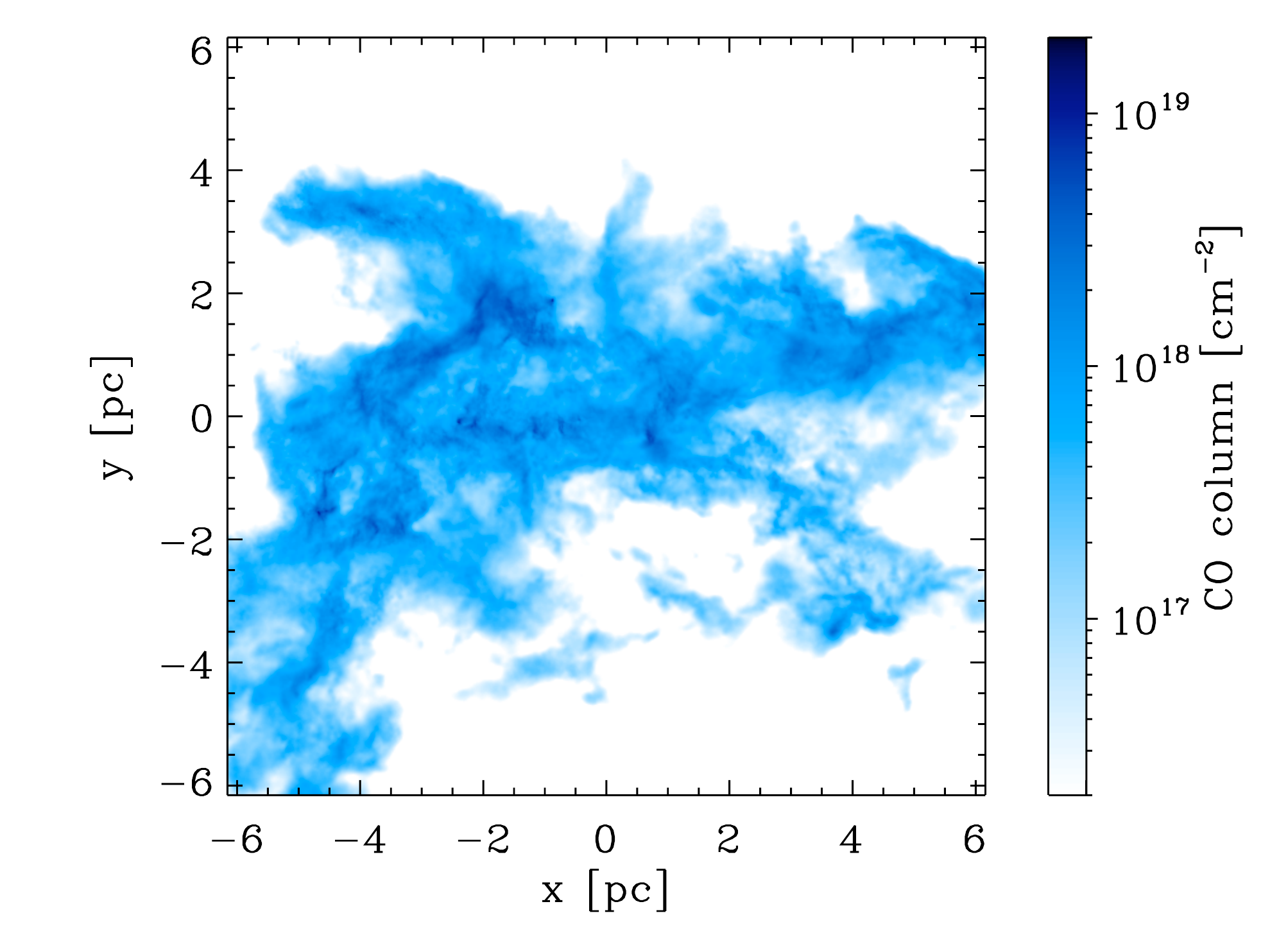}
\includegraphics[width=3.6in]{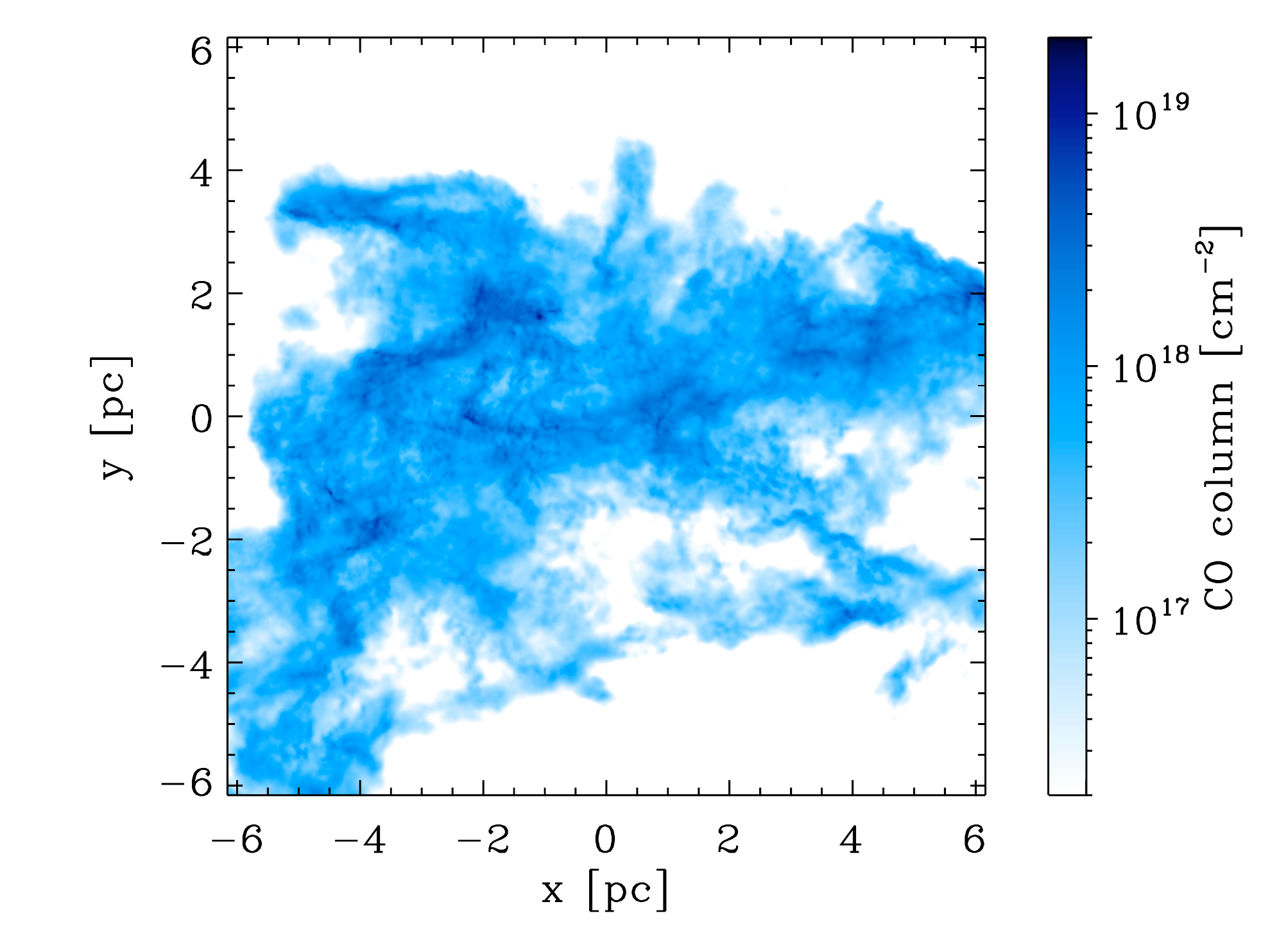}
}
\caption{CO column density projection at $t = 2.2 \: {\rm Myr}$ in run D1 (left-hand panel) and 
run D2 (right-hand panel). More small-scale, low column density structure is visible in run D2,
which started with its hydrogen in molecular form, than in run D1, which started with its hydrogen
in atomic form. However, the high column density structures are quite similar in both cases. 
\label{fig:cocol}}
\end{figure*}

In Figure~\ref{rhoco}, we show how the fractional abundance of CO, $x_{\rm CO} \equiv n_{\rm CO} / n$,
varies as a function of $n$ in runs A, D1 and D2. Results are shown at the same output times as in
Figure~\ref{rhoh2}. No results are plotted for runs B or C, because $x_{\rm CO} = 0$ by design in these
runs. Several points should be noted. First, the CO abundance in run A is tiny, as is to be expected given
the low H$_{2}$ abundance and the absence of dust shielding. Second, the distribution of CO abundances
in runs D1 and D2 are very similar, despite the difference in the initial conditions for these two runs. This
suggests that the CO content of the gas is primarily sensitive to its {\em current} H$_{2}$ content and 
density structure. Any sensitivity to the chemical history of the gas, which is very different in runs D1 and
D2, must be small. Additional support for this point is provided by Figure~\ref{fig:cocol}, which shows a
CO column density projection of the clouds in runs D1 and D2 at $t = 2.2 \: {\rm Myr}$. Although there
appears to be somewhat more structure at low CO column densities in run D2 than in run D1, the main
features of the cloud are very similar in both simulations, and there is no clear observational signature
of the different chemical histories that one could point to in these runs.

\subsubsection{Dominant thermal processes}
\label{dom_therm}
\begin{figure*}
\includegraphics[width=5.0in]{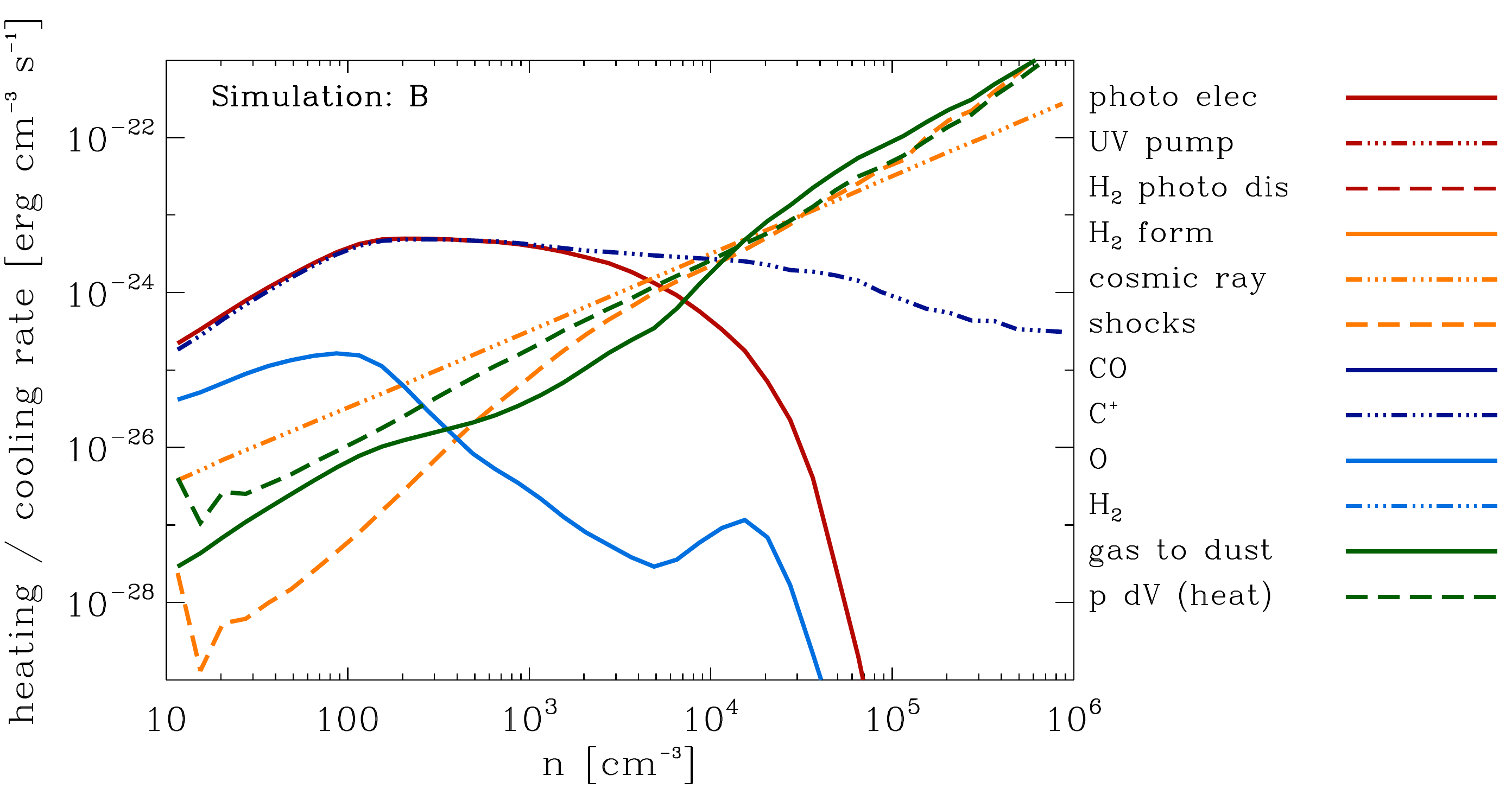}
\caption{Median heating and cooling rates per unit volume in run B, 
plotted as a function of the hydrogen nuclei number density $n$,  at a time just before the onset of star formation.
\label{heatcool_atom}}
\end{figure*}

\begin{figure*}
\includegraphics[width=5.0in]{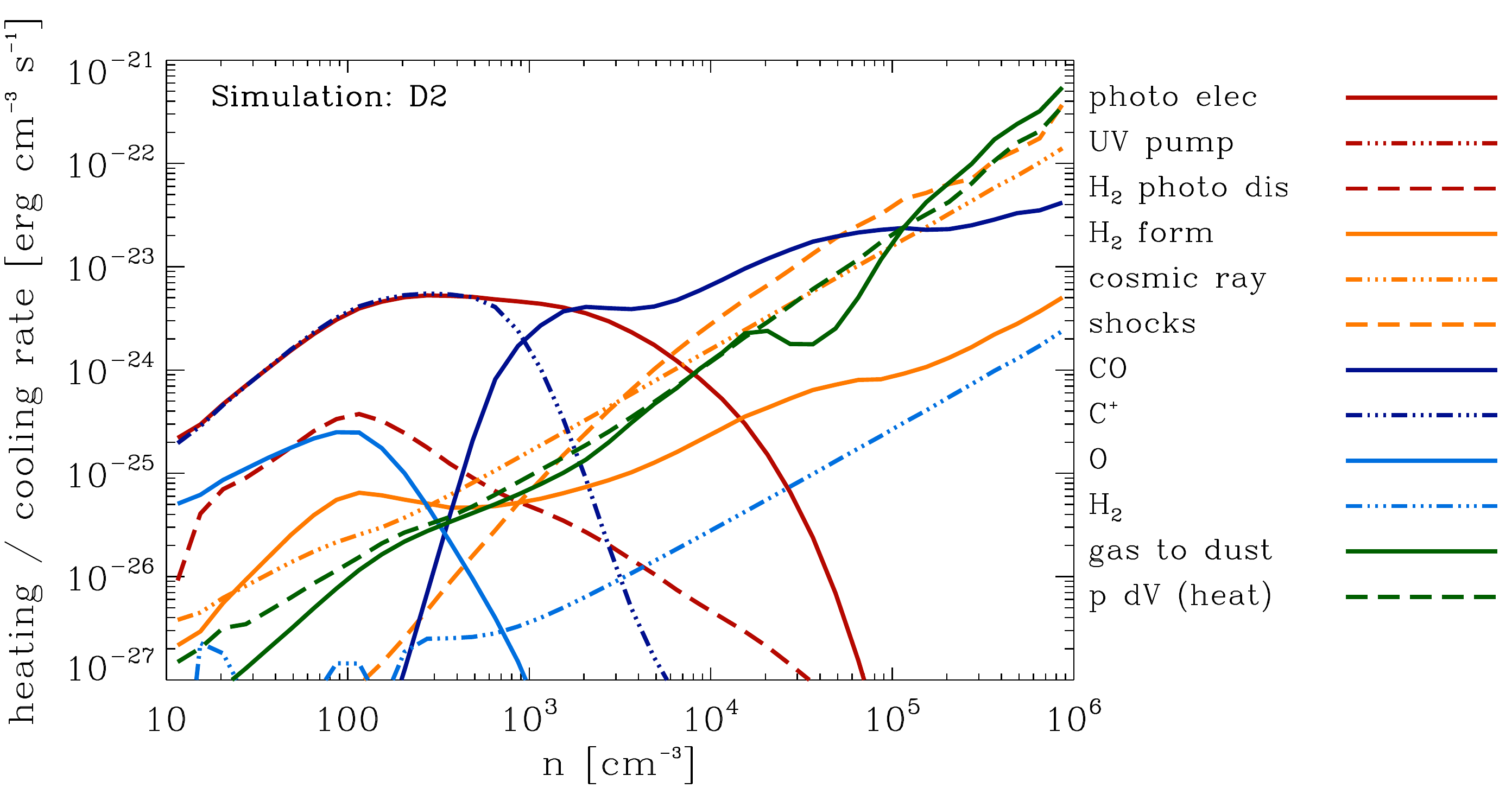}
\caption{As Figure~\ref{heatcool_atom}, but for run D2 \label{heatcool_mol}}
\end{figure*}

To help us to understand why the presence or absence of molecules appears to have such a limited effect
on the temperature structure of the clouds and the progress of star formation within them, it is useful to examine
how the dominant heating and cooling process vary with density within an atomic cloud and a molecular cloud.
We have therefore calculated the median heating/cooling rate as a function of density for each process 
included in our thermal model for runs B and D2 at a time just before the onset of star formation.
The results are plotted in 
Figure~\ref{heatcool_atom} (for run B) and Figure~\ref{heatcool_mol} (for run D2). Values are plotted only for
the most important processes; minor contributions to the cooling from e.g.\ H$^{+}$ recombination or the collisional
ionization of atomic hydrogen are omitted. We have chosen to plot the median values rather than the mean
values because the latter are strongly influenced by dissipation in shocks, which occupy a relatively small fraction
of the simulation volume but dissipate a significant amount of energy within the shocked regions 
\citep[see e.g.][]{pp09}. The median values give a fairer picture of the balance of heating and cooling rates in the 
bulk of the gas.

Figure~\ref{heatcool_atom} demonstrates that in the atomic cloud, there are three main regimes. At densities 
$n < 2000 \: {\rm cm^{-3}}$, the heating of the gas is dominated by photoelectric emission from dust grains, and
the cooling is dominated by fine structure emission from C$^{+}$. Between $n \sim 2000 \: {\rm cm^{-3}}$ and
$n \sim 2 \times 10^{4} \: {\rm cm^{-3}}$, C$^{+}$ remains the main coolant, but photoelectric heating becomes
much less effective, owing to the increasing visual extinction of the cloud at these densities. In this density range,
adiabatic compression of the gas (indicated in the plot as the pdV term), dissipation of turbulent kinetic energy
in shocks, and cosmic ray ionization heating are all significant sources of heat.
Finally, at densities greater than $n \sim 2 \times 10^{4} \: {\rm cm^{-3}}$, the timescale for energy transfer
between gas and dust becomes short enough to couple the gas and dust temperatures together, at which 
point this (followed by thermal emission from the dust grains) becomes the most important cooling process.
Weak shocks and adiabatic compressions together dominate the heating of the gas in this regime, each
contributing close to half of the total heating rate. 

Figure~\ref{heatcool_mol} shows that in the molecular cloud, a number of additional processes come
into play: H$_{2}$ and CO provide additional cooling channels through their rotational line emission,
but the presences of H$_{2}$ also introduces additional heating processes such as 
H$_{2}$ photodissocation heating \citep{bd77}, or heating due to the pumping of highly excited vibrational
levels of H$_{2}$ by UV photons \citep{bht90}. Nevertheless, despite the additional complexity, we can
again identify three main regimes marked out by different dominant processes. At $n < 1000 \: {\rm cm^{-3}}$,
the behaviour is very similar to that in the atomic run: C$^{+}$ is the dominant coolant, while most of the
heating comes from photoelectric emission from dust. Above $n = 1000 \: {\rm cm^{-3}}$, C$^{+}$
quickly gives way to CO as the dominant coolant, reflecting the fact that the gas becomes CO-dominated
at around this density (see Figure~\ref{rhoco}), and the photoelectric heating rate also begins to fall off
with increasing density. The fact that these changes occur at a very similar density is no coincidence: 
the photoelectric heating rate and the CO photodissociation rate have a very similar dependence on
the visual extinction of the gas, and so both become unimportant at roughly the same point.
CO remains the dominant coolant between $n = 1000 \: {\rm cm^{-3}}$ and $n \sim 10^{5} \: {\rm cm^{-3}}$,
but photoelectric heating quickly becomes irrelevant, and dissipation in shocks becomes the main
source of heat.  Finally, at $n > 10^{5} \: {\rm cm^{-3}}$, dust takes over from CO as the most important
coolant, and pdV heating becomes almost as important as shock heating.

Figure~\ref{heatcool_mol} also illustrates that cooling by H$_{2}$ is never particularly important:
at best, it contributes only a few percent of the total cooling rate, and at most densities contributes far
less than this. In addition, it demonstrates that H$_{2}$ formation heating is unimportant in run D2,
which is unsurprising given the fully molecular initial conditions used for this run. A similar plot for
run D1 would show a much larger contribution from H$_{2}$ formation heating at densities between
$n = 10^{3} \: {\rm cm^{-3}}$ and $n = 10^{4} \: {\rm cm^{-3}}$.

If we compare Figures~\ref{heatcool_atom} and \ref{heatcool_mol}, we can see why the presence of
H$_{2}$ and CO appears to have such a limited effect on the behaviour of the cloud. Below $n = 
1000 \: {\rm cm^{-3}}$ (which is, let us not forget, more than three times higher than the initial mean
density of the gas), the cooling is dominated by C$^{+}$ in both simulations, and so H$_{2}$ and
CO have very little influence on the thermal behaviour of the gas. At higher densities, CO takes
over from C$^{+}$ as the dominant coolant in run D2. However, if we compare the CO cooling rate
at e.g.\ $n = 10^{4} \: {\rm cm^{-3}}$ in run D2 with the C$^{+}$ cooling rate at the same density in
run B, we see that they are surprisingly similar -- the C$^{+}$ cooling rate is smaller than the CO
cooling rate, but only by a factor of 2--3, despite the much larger energy required to excite the
C$^{+}$ line compared to the CO $J = 1 \rightarrow 0$ line. It is this fact that ultimately renders
the CO inessential: if we were to remove it, the temperature of the gas would rise slightly, and the
cooling rate would fall by a factor of a few, but otherwise, the behaviour of the cloud would hardly
change, and hence its ability to form stars would not be greatly affected.  Since, as we have
already seen, H$_{2}$ cooling is also unimportant, we can therefore conclude that molecular
cooling in general is not required in order to form stars, given the initial conditions examined here.
In a realistic model, the gas that forms stars will be molecular (as in runs C, D1 and D2), but even if 
it were not (as in run B), it would still form stars at almost the same rate.

Finally, we note that although the chemical model used in our simulations does not include atomic
carbon, which could potentially be an important low-temperature coolant, we would not expect its
inclusion to significantly affect the star-formation rate. Atomic carbon is unlikely to provide more
cooling at low temperatures than CO, and as we have already seen, the star formation rate is
insensitive to the amount of CO present in the gas, implying that the star formation rate will also
be insensitive to the ratio of C$^{+}$ to C. We have tested this assumption by resimulating the
cloud from run D2 using a more detailed chemical model that does include atomic carbon
(the NL99 model described in 
\citealt{gc11}, which is a slightly modified version of a model first introduced by \citealt{nl99}).
As expected, we find that the star formation rate in this run is 
essentially the same as  in run D2.

\subsection{Influence of numerical resolution}
\label{resn}
\begin{figure}
\includegraphics[width=3.6in]{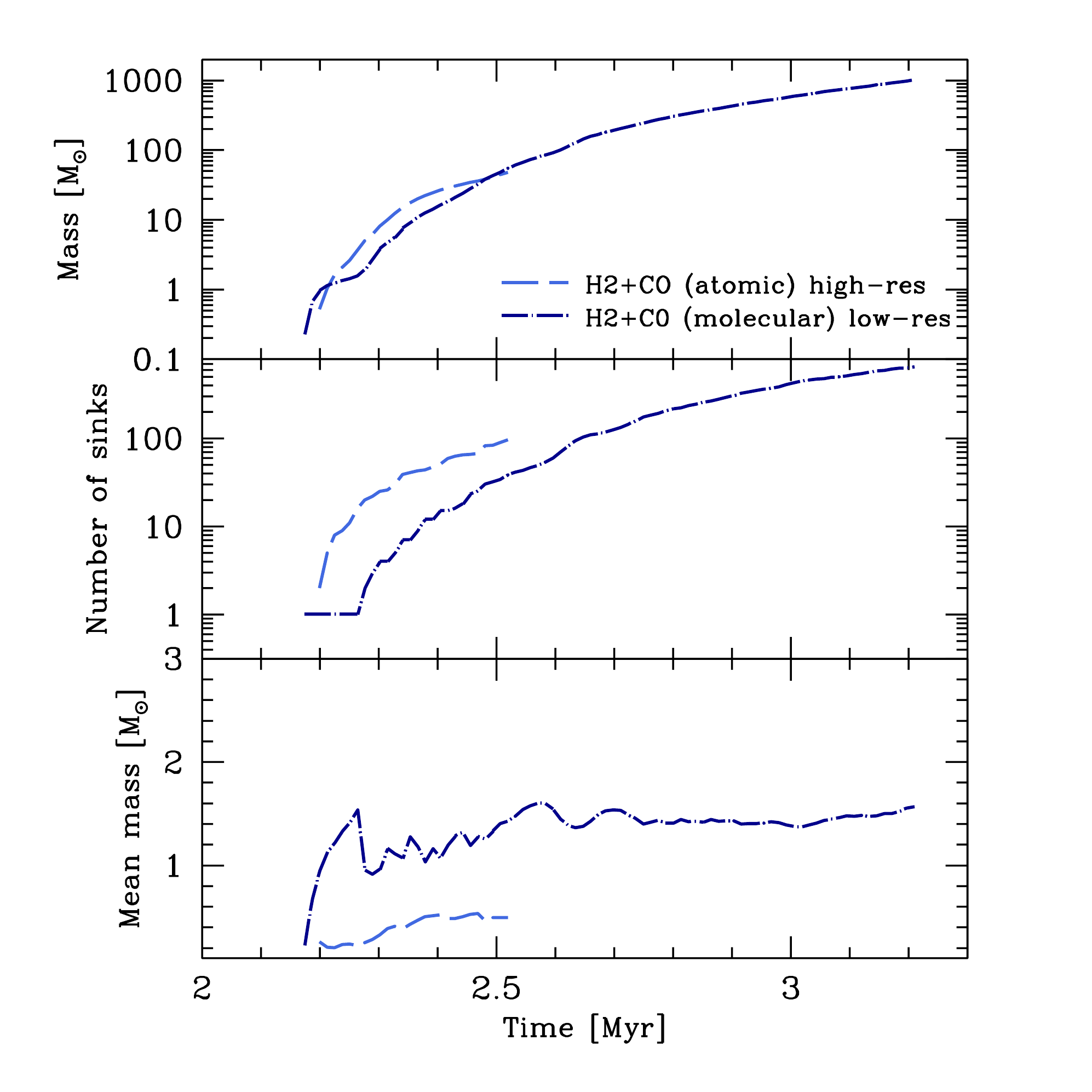}
\caption{{\em Upper panel:} mass in sinks as a function of time in runs D2 (dot-dashed line)
and D3 (dashed line). These two runs
use the same chemical model and initial conditions, and differ only in numerical resolution, with run D3 having
ten times better mass resolution than run D2.  {\em Middle panel:} number of sinks formed in each simulation,
plotted as a function of time. {\em Lower panel:} mean sink mass versus time in the same four simulations. For
reference, the mass resolution in run D2 was $0.5 \: {\rm M_{\odot}}$, while in run D3 it was $0.05 \: {\rm M_{\odot}}$.
\label{SF-hist-highres}}
\end{figure}

To establish the extent to which our results are affected by the limited mass resolution of our 
simulations, we performed a single simulation with the same initial conditions and chemical
model as run D2, but which had ten times more SPH particles, and hence a mass resolution
of $0.05 \: {\rm M_{\odot}}$, ten times smaller than in our standard runs. We denote this run 
as D3. In Figure~\ref{SF-hist-highres}, we compare the growth of the mass incorporated into sink 
particles, the number of sink particles formed, and the mean sink particle mass in runs D2 and D3. 
We see that the high resolution run D3 begins to form stars at almost the same point as the lower
resolution run D2, and that the stellar mass increases at very nearly the same rate, although the
much higher computational cost of the higher resolution run means that we are unable to follow
the evolution of the system for as long as in the low resolution run. We also note the high resolution
run forms a significantly larger number of stars than the low resolution run, with a mean stellar mass
that is a factor of a few smaller.

These results suggest that the rate at which stars form within our model clouds is not strongly influenced
by the limited mass resolution of our simulations. This is what one would expect if the main factor determining
the star formation rate is the formation of self-gravitating, dense cores \citep[see e.g.][]{km05,pn11}, 
as most of these cores are well resolved even in our low resolution simulations. The mass distribution of the stars 
that form, on the other hand, is sensitive to fragmentation occurring within the cores, which is not well resolved within our
low resolution models, as the differences in the lower two panels of Figure~\ref{SF-hist-highres} clearly 
demonstrate.

We therefore conclude that our major results, which concern the timing and rate of star formation within
the different clouds, should be insensitive to our choice of numerical resolution, but that higher resolution
simulations will be needed to fully explore the sensitivity of the stellar initial mass function to the chemical
make-up of the gas.

\section{Discussion}
\label{discuss}
Our simulations demonstrate that once a dense, gravitationally unstable cloud of gas has formed in the ISM,
molecular cooling is not required in order to convert this gas into stars. In the absence of molecules, the cooling
provided by the C$^{+}$ fine structure line at $n < 2 \times 10^{4} \: {\rm cm^{-3}}$ and by dust at $n > 2 \times
10^{4} \: {\rm cm^{-3}}$
are sufficient to allow stars to form. However, we cannot conclude solely from this that molecular gas is
an inessential component of the star formation process. A skeptic could well argue that molecular cooling may
be required in order to form the dense cloud of gas that we take as the initial conditions for our simulations, 
an assertion that our current results can obviously do nothing to address. Fortunately, the issue of the assembly
of cold, dense clouds of gas from the warm, neutral ISM has already been investigated by a number of authors
\citep[see e.g.][]{ki02,ah05,vs07,h08,hh08,banerjee09}.

These studies typically make use of a setup that involves converging flows of warm atomic gas, with an initial
temperature and density corresponding to the warm neutral medium (WNM). The thermal models adopted vary
somewhat in complexity, from a simple two-parameter cooling function introduced by \citet{ki02} and used in
a number of later works \citep[e.g.][]{vs07,banerjee09}, to the more detailed treatment used in e.g.\ 
\citet{ah05} that accounts explicitly for photoelectric heating from dust and fine structure cooling from C$^{+}$ and O.
However, one common feature is that these models do not account for cooling coming from H$_{2}$ or CO.
Indeed, they typically do not include any non-equilibrium chemistry. Despite this, these models readily produce
cold, dense clouds of gas in the region where the converging flows intersect. These cold clouds are produced
initially by a thermal instability acting in the WNM that relies primarily on C$^{+}$ cooling. The clouds can then 
collide and agglomerate, forming gravitationally bound structures. 

The typical densities and masses of the cold clouds formed in these converging flow studies span a wide range
of values, depending on the size of the simulation volume and the effective resolution of the simulation.
Large-scale converging flow simulations, such as that performed by  \citet{banerjee09}, have shown themselves 
capable of producing cold clouds with properties similar to those we take as our initial conditions (i.e.\ a mean density 
of a few hundred ${\rm cm^{-3}}$ and a cold gas mass of $10^{4} \: {\rm M_{\odot}}$) even though these simulations
do not include molecular cooling. Models of cloud formation on the scale of the Galactic disk also appear to have
no difficulty in producing large cold gas clouds without needing to invoke molecular cooling \citep[see e.g.][]{tt09}. 
Taken together with our results, this implies that molecular gas is not necessary for star formation: it is possible to form
both cold, dense clouds, and also stars within these clouds, without needing to rely on molecular cooling. 

If molecular gas is not a prerequisite for star formation, then why is it that we find such a good observational 
correlation between the surface density of molecular gas, and the star formation rate surface density? An
obvious explanation is that both the presence of molecular gas and the propensity of a gas cloud to form
stars are correlated with some third factor, such as the column density of the cloud. An argument along these
lines was given by  \citet{schaye04}, who pointed out that in regions of the Galactic disk with gas surface densities
of less than $10 \: {\rm M_{\odot}} \: {\rm pc^{-2}}$, ionization and heating of the gas by the extragalactic radiation
field prevent the formation of a cold, neutral phase. The transition from the warm neutral medium to the cold neutral
medium is associated with a dramatic decrease in the Jeans mass of the gas, and \citet{schaye04} argues that
this is the trigger for star formation. The transition from atomic to molecular gas coincidentally occurs at roughly
the same surface  density \citep[see e.g.][]{brown03,kmt09}, and so the result is a correlation between the presence 
of molecular gas and the formation of stars.

The \citet{schaye04} model assumes that all cold clouds form stars, but as we have seen in this paper, this
is probably an oversimplification. In a more recent paper, \citet{klm11} look in more detail at the chemistry
and thermodynamics of cold clouds in the ISM. They use simple 1D cloud models that assume chemical and 
thermal equilibrium to explore a range of different cloud densities and visual extinctions, and 
show that in these models, the transition from atomic to molecular hydrogen is well correlated with a further
sharp drop in the equilibrium gas temperature. This correlation is not a result of H$_{2}$
cooling. Rather, it occurs because the conditions required in order to attain a low gas temperature -- high
densities to boost C$^{+}$ cooling, and efficient dust shielding to suppress photoelectric heating
-- are similar to those required to produce a high equilibrium H$_{2}$ fraction. They then argue that 
star formation is strongly correlated with regions of cold gas, owing to the $T^{3/2}$ temperature dependence
of the Bonnor-Ebert mass scale, which makes gravitational fragmentation much easier to bring about in cold
gas than in warm gas. In both of these models, the correlation between H$_{2}$ and star formation  comes about because
the H$_{2}$ traces (but does not cause) the regions where the thermal pressure is low enough to allow stars
to form.

The results from our present study provide strong support for this picture. They show that H$_{2}$  cooling
plays an insignificant role in determining the cloud temperature, that the differences between the temperatures
of clouds cooled solely by C$^{+}$ or by a mix of C$^{+}$ and CO are very similar, and that the key factor 
enabling star formation within the clouds is the shielding of the interstellar radiation field by dust. In the absence
of this shielding, the temperature of the gas remains high, star formation is strongly suppressed, and the H$_{2}$ 
abundance is very small.

\section{Conclusions}
\label{conc}
In this study, we have investigated whether or not the formation of molecular gas is a prerequisite for
star formation. We have performed simulations using several different chemical models: one in which
the gas is assumed to remain atomic throughout, a second in which H$_{2}$ formation is included, but
CO formation is not, and a third which follows both H$_{2}$ and CO formation. We find only minor 
differences in the nature and rate of star formation in these simulations. In contrast, disabling the 
effects of dust shielding has a very strong effect: the gas temperature does not fall much below 100~K
and the formation of stars is strongly suppressed. 

We infer from these results that the observational correlation between H$_{2}$ and star formation is
not a causal relationship: H$_{2}$ and CO are not required for star formation, and the fact that we
find a good correlation between the H$_{2}$ surface density and the star formation rate surface density
simply reflects the fact that both are correlated with some third factor. Our results suggest that the
key factor is the ability of the clouds to shield themselves effectively against the interstellar radiation
field: clouds that are too diffuse to shield themselves do not cool and hence form few, if any stars.
Since effective shielding of the UV background is also required in order to form large abundances
of H$_{2}$ or CO \citep{g10,gm11}, this naturally leads to a correlation between molecular gas and star
formation without the necessity for a direct causal link between them. Our current results do not allow 
us to establish whether the column density of dust required to provide effective shielding is independent 
of the cloud properties, or a function of the mean density of the gas; this will be addressed in future work.

Finally, it is interesting to speculate about the possible observational consequences of this result. One
possible way of testing this model would be to look for regions in which ongoing star formation is not
accompanied by significant amounts of molecular gas. If molecular gas is required for star
formation, then it would be very difficult to explain the existence of such regions, whereas they are
easily accommodated within our  model. Unfortunately, such regions are likely to be rare, since they
require the molecular gas fraction to be far out of equilibrium, and as our models show, high molecular
fractions can be reached within a single free-fall given typical Galactic conditions. If our results hold
at lower metallicities (which is a reasonable assumption, but one which must be tested by future 
simulations), then one promising place to look for such regions would be
within low metallicity star-forming dwarf galaxies such as I~Zw~18 or
DDO~154, since in these systems the characteristic chemical timescales will be much longer. Indeed, efforts 
to date to detect CO within these galaxies have been unsuccessful \citep{leroy07,komugi11}, although whether 
this is due to a general deficit of molecular gas or simply a low CO-to-H$_{2}$ ratio remains unclear. 

\section*{Acknowledgements}
The authors would like to thank M.-M.\ {Mac Low} and M.\ Krumholz for 
stimulating discussions regarding the role that molecular gas plays in
star formation. They also thank J.\ Schaye and the referee, J.\ Black,
for their feedback on an earlier draft of this paper.
The authors acknowledge financial support from the Landesstiftung
Baden-W\"urrtemberg via their program International Collaboration II
(grant P-LS-SPII/18), from the German Bundesministerium f\"ur
Bildung und Forschung via the ASTRONET project STAR FORMAT (grant
05A09VHA), from the DFG under grants no.\  KL1358/4 and KL1358/5,
and from a Frontier grant of Heidelberg University sponsored by the German
Excellence Initiative. The simulations reported on in this paper were
primarily performed using the {\em Kolob} cluster at the University of
Heidelberg, which is funded in part by the DFG via Emmy-Noether grant
BA 3706, and the authors would like to thank R.\ Banerjee, I.\ Berentzen,
P.\ Girichdis and G.\ Marcus for the hard work that they have done to keep 
{\em Kolob} functioning.

\end{document}